\documentclass[aps,prl,twocolumn,showpacs,groupedaddress]{revtex4}

\usepackage{graphicx}
\usepackage{amsmath}
\usepackage{color}
\usepackage{epsfig}
\usepackage{epstopdf}

\begin{document}

\title{Time-Resolved Upconversion of Entangled Photon Pairs}

\author{Kevin A. O'Donnell$^{1}$ and Alfred B. U'Ren$^{2}$ } 
\affiliation{$^1$Divisi\'on de F\'isica Aplicada, Centro de Investigaci\'on Cient\'ifica y de Educaci\'on Superior de Ensenada, Km.~107 Carretera Tijuana-Ensenada, Ensenada, B.C. 22860, M\'exico\\
$^2$Instituto de Ciencias Nucleares, Universidad Nacional Autonoma de M\'exico, Apdo. Postal 70-543, M\'exico D.F.  04510}

\date{\today}

\begin{abstract}
In the process of spontaneous parametric downconversion, photons from a pump field are converted to signal and idler photon pairs in a nonlinear crystal.  The reversed process, or upconversion of these pairs back to single photons in a second crystal, is also possible. Here, we present experimental measurements of the upconversion rate with a controlled time delay introduced between the signal and idler photons. As a function of delay, this rate presents a full width at half maximum of $27.9 \, \mathrm{fs}$ under our experimental conditions, and we further demonstrate that group delay dispersion of the photon pairs broadens this width. These observations are in close agreement with our calculations, thus demonstrating an ultrafast, nonclassical correlation between the signal and idler waves. 
\end{abstract}


\maketitle

There are few ways of characterizing photon wavepackets on an ultrafast time scale.  While temporal resolution of a few fs can be necessary, optical detectors have a response that is slower by, at least, several orders of magnitude.  One way of overcoming this limitation is with two-photon interference, as pointed out by Hong, Ou, and Mandel\cite{HOMI87}.  With fs-level resolution, they determined the distribution of arrival time intervals between the signal and idler photons produced by spontaneous parametric downconversion (SPDC).  Refinements of this approach have since measured relative signal/idler group delays to a remarkable precision of $0.1 \, \mathrm{fs} \, $\cite{Brann00}.

Other interactions produced with nonclassical light have been considered in previous works\cite{bunch, Abram86, Dayan05, Dayan07, Harris07}.  These interactions inspire another possible ultrafast characterization method: to introduce a controlled delay between the signal and idler photons of SPDC, overlap them spatially in a nonlinear crystal, and detect the sum-frequency photon produced should they upconvert.  This approach would yield high-resolution temporal information about the original quanta; indeed, it is analogous to the widely-utilized autocorrelation method of characterizing classical laser pulses\cite{Diels06}, where fs-level resolution is common.  Moreover, it will be shown here that such an approach is dispersion-sensitive, in contrast to the Hong-Ou-Mandel interferometer\cite{Stein92}.  Some years ago, this type of time-resolved upconversion was studied with high-gain ($\approx \! 10^{10}$) parametric light\cite{Abram86}.  However, it was soon pointed out that the approach would be of no use at the photon level\cite{HOMI87}, essentially due to the weak nonlinearities of crystals then available.

Present-day crystals have stronger nonlinearities and it is therefore time to readdress this issue.  Using an unusually bright SPDC source, Dayan \textit{et al.}\cite{Dayan05} have provided the first observations of signal/idler paired upconversion in a nonlinear crystal, where photon pair members were kept perfectly synchronized.  Yet, two recent theoretical works have considered time-resolved, pairwise upconversion\cite{Dayan07, Harris07}, and related experimental results have been reported\cite{Peer05} that we will discuss later.  In part, our purpose is to present novel, time-resolved experimental results, but we begin with theoretical considerations.

The two-photon state of SPDC may be written as 
\begin{equation}
|\Psi_{DC}\rangle= | 0 \rangle+\eta \int d^3 \vec{k}_s \int d^3 \vec{k}_i f_{DC}(\vec{k}_s,\vec{k}_i) \hat{a}^\dag_s (\vec{k}_s) \hat{a}^\dag_i(\vec{k}_i) | 0 \rangle, \label{E:state} 
\end{equation}
where $\eta$ is related to the conversion efficiency, $f_{DC}(\vec{k}_s,\vec{k}_i)$ is the joint amplitude, $a_\mu^\dagger(\vec{k}_\mu)$ is the creation operator for the signal ($\mu \! = \! s$) or idler ($ \! \mu= \! i$) modes with wavevector $\vec{k}_\mu$, and $|0\rangle$ is the vacuum state. We want to describe the upconversion of these photon pairs into single photons at the sum frequency.  In the spontaneous conversion limit, the upconverted state $|\Psi_{UC}\rangle$ can be expressed in the form
\begin{equation}
|\Psi_{UC}\rangle=(1+\hat{U_1})|\Psi_{DC}\rangle=\left(1+\frac{1}{i \hbar} \int \limits_{0}^{t} d t' t' \hat{H}(t') \right)|\Psi_{DC}\rangle, \label{Unitary}
\end{equation} 
\noindent where $(1+\hat{U}_1)$ is the unitary time-evolution operator. 
The upconversion Hamiltonian $\hat{H}(t)$ can be expressed in terms of the nonlinearity $d(\vec{r})$ and electric field operators $\hat{E}_\mu^{(\pm)}(\vec{r},t)$ as 
\begin{equation}
\hat{H}(t)=\int d^3 \vec{r}\ \ d(\vec{r}) \, \hat{E}_u^{(-)}(\vec{r},t) \, \hat{E}_s^{(+)}(\vec{r},t) \, \hat{E}_i^{(+)}(\vec{r},t) + \mathrm{H. C.} \, , \label{E:Hamiltonian} \end{equation}
\noindent where $ \! \mu= \! u$ now also includes the upconverted mode, {\footnotesize \rm{+/$-$}} denotes positive-/negative-frequency parts, and the second term is the hermitian conjugate. By expressing each of the electric fields as a Fourier integral in $\vec{k}_\mu$, and performing the integral of Eq.~(\ref{Unitary}), we obtain
\begin{eqnarray}
\hat{U}_1&=&\frac{2 \pi}{i \hbar} \int d^3 \vec{k}_u \int d^3 \vec{k}_s \int d^3\vec{k}_i \, f^{*}_{UC}(\vec{k}_s,\vec{k}_i,\vec{k}_u) \nonumber \\ &\times& \delta(\omega_u-\omega_s-\omega_i) \, \hat{a}^\dag_u(\vec{k}_u) \, \hat{a}_s(\vec{k}_s) \, \hat{a}_i(\vec{k}_i) \, . \end{eqnarray} 
Defined in terms of $k_{+\nu} \! = \! k_{s \nu} \! + \! k_{i \nu}$ (with $\nu \! = \! x,y$), $f_{UC}(\vec{k}_s,\vec{k}_i,\vec{k}_u)$ follows from the integral of Eq.~\ref{E:Hamiltonian} as 
\begin{equation}
f_{UC}(\vec{k}_s,\vec{k}_i,\vec{k}_u)=\delta(k_{+x}-k_{ux})\delta(k_{+y}-k_{uy}) \Phi_{UC}(\vec{k}_s,\vec{k}_i,\vec{k}_u) \end{equation} 
\noindent where $\Phi_{UC}(\vec{k}_s,\vec{k}_i,\vec{k}_u)$ is given by $\mbox{sinc}(\beta) \exp(-i \beta)$ with $\beta= \Delta k_z L/2$ and $\Delta k_z=k_{pz}-k_{sz}-k_{iz}-k_g$, $L$ is the upconversion crystal length, and $k_g=2 \pi/\Lambda$ is the wavenumber associated with the crystal poling period $\Lambda$. 
We now assume that the SPDC pump is monochromatic with frequency $\omega_p$. In this case we can write $f_{DC}(\vec{k}_s,\vec{k}_i)$ as $\delta(\omega_s \! + \! \omega_i \! - \! \omega_p) \, g(\vec{k}_s,\vec{k}_i)$ and the upconverted state  of Eq.~(\ref{Unitary}) becomes $|\Psi_{UC}\rangle=|0\rangle+\kappa \int d^2\vec{k}^\bot_u F(\vec{k}_u^\bot)a_u^\dagger(\vec{k}_u^\bot,\omega_p)| 0 \rangle$. This integral is over only transverse components of $\vec{k}_u$, the creation operator label is now expressed in terms of $\vec{k}_u^\bot$ and  $\omega_p$, $\kappa$ includes various constants, and $F(\vec{k}_u^\bot)$ is given by 
\begin{eqnarray}
F(\vec{k}_u^\bot)&=&\int d^3 \vec{k}_s \Phi_{UC}^*(\omega_s,\vec{k}_s^\bot;\omega_p-\omega_s,\vec{k}_i^\bot; \omega_p,\vec{k}_u^\bot)\nonumber \\ &\times& g(\omega_s,\vec{k}_s^\bot;\omega_p-\omega_s,\vec{k}_u^\bot) \, , \end{eqnarray} 
where both $\Phi_{UC}(\cdot)$ and $g(\cdot)$ are now written in terms of the new variables.

For a plane-wave SPDC pump, $g(\omega_s,\vec{k}_s^\bot;\omega_p-\omega_s,\vec{k}_u^\bot)$ is given as $\delta(k_{+x}) \, \delta(k_{+y})  \, \Phi_{DC}(\vec{k}_s,\vec{k}_i)$ where $\Phi_{DC}(\vec{k}_s,\vec{k}_i)=\mbox{sinc}(\beta) \exp(i \beta)$, and we have assumed that both crystals are characterized by the same $\beta$.  We further assume that, before the upconversion crystal, the signal and idler photons propagate through dispersive optical elements giving each a spectral phase $\exp[i \phi_\mu(\omega_\mu)]$, that they encounter a limiting aperture described by pupil function $P(\vec{k_\mu})$, and that the signal mode is delayed by $\tau$ (these operations amount to multiplying $f_{DC}(\vec{k}_s,\vec{k}_i)$ of Eq.~(\ref{E:state}) by appropriate factors, which simply carry through the analysis above). The upconverted state may then be written as $|\Psi_{UC}\rangle=|0\rangle+\kappa^\prime|\Psi_{UC}^{(1)}\rangle$ with 
\begin{equation}
|\Psi_{UC}^{(1)}\rangle = \left[\int d \omega_s S(\omega_s) e^{-i \omega_s \tau}\right] \hat{a}_u^\dagger(0,\omega_p)\, |0\rangle \, , \label{E:PsiPUC}
\end{equation}
\noindent where
\begin{equation}
S(\omega_s)=\int d^2\vec{k}_s^\bot \mbox{sinc}^2 (\overline{\beta}) e^{i \overline{\Delta k}_z L} e^{i[\phi_s(\omega_s)+\phi_i({\overline{\omega}_i})]} P(\vec{k}_s) P(\overline{\vec{k}}_i)  , \label{E:Somega} \end{equation}
and overlined quantities are evaluated with $\omega_i \! = \! \omega_p \! - \! \omega_s$ and $\vec{k}_i^\bot \! = \! -\vec{k}_s^\bot$.

Our main theoretical results are Eqs.~(\ref{E:PsiPUC}-\ref{E:Somega}).  Here, $S(\omega_s)$ is related to the amplitude spectrum of signal-mode photons that participate in the upconversion process.  Still, it is also clear that $S(\omega_s)$ depends on the dispersion imposed on \textit{both} the signal and idler modes through the second  exponential factor of Eq.~(\ref{E:Somega}).  At the least, this indicates that control of dispersion is necessary in the experimental work to be described.  We also note that previous works\cite{Dayan07, Harris07} have derived what amount to axial limits of Eqs.~(\ref{E:PsiPUC}-\ref{E:Somega}),  and for us it is also straightforward to obtain a simplification of  Eq.~(\ref{E:Somega}) in the axial case.  However, the factors related to phasematching are strongly angle-dependent  in cases later considered here and, to obtain accurate results, retaining the form of Eq.~(\ref{E:Somega}) is necessary.

The experiment is shown in Fig.~1.  The pump laser was a single-mode, continuous-wave ring laser of wavelength $532 \, \mathrm{nm}$ and power $1 \, \mathrm{W}$.  Its beam was focused to a $45 \, \mu \mathrm{m}$  waist in an uncoated, magnesium oxide doped, $5 \, \mathrm{mm}$ long lithium niobate crystal, which produced the down-converted light.  The crystal was periodically-poled and temperature-controlled to permit phasematching to nearly axial, co-polarized, frequency-degenerate photon pairs.  The emission within a cone of half-angle $2^\circ$ was accepted by the aperture of the optical system that followed.

%1
\begin{figure}
\includegraphics[width=0.48 \textwidth]{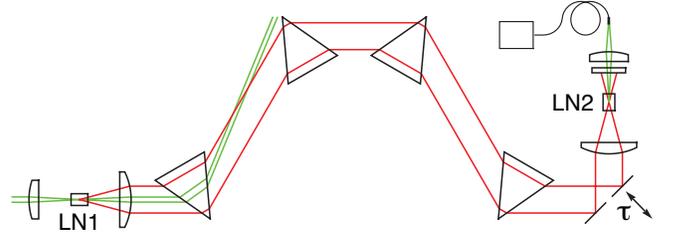}
\caption{ Experimental apparatus.  Parametric downconversion from lithium niobate crystal LN1 is collimated and sent through a series of 4 Brewster prisms, to be upconverted in a similar crystal LN2.  Two mirrors (one fixed, the other on a piezoelectric stage, translated as shown) introduce a time difference $\tau$ between the upper and lower optical paths.} 
\end{figure}

The light was collimated by a $75 \, \mathrm{mm}$ focal length lens and was sent through a SF14 Brewster prism, which deviated the pump beam from the system.  The downconverted light passed through four such prisms, symmetrically arranged, to compensate for the chirp of the photon wavepackets arising from all dispersive media in their path\cite{Dayan05}.  The tip-to-tip spacing of the first or second prism pair was $352 \, \mathrm{mm}$.

In the collimated downconverted beam, transverse momentum conservation implies that a given photon will have its pair member on the opposite side of beam center. Thus, in the light exiting the prisms of Fig.~1, reflecting the upper part of the beam (containing, say, signal photons) from a displaceable mirror and reflecting the lower part (idler photons) from a second, fixed mirror provided the signal/idler time difference $\tau$.  At beam center, a $1.5 \, \mathrm{mm}$ intermirror gap guaranteed that any signal or idler photon could interact with only one mirror, thus dividing reflected pair members into two distinct, time-shifted modes.  Angular mirror co-alignment was achieved with a subsidiary Michelson interferometer (not shown), where the mirrors acted as a single end-mirror.   There, a tunable source was used to find the translation setting that made the mirrors coplanar ($\tau \! = \! 0$), for which fringes remained aligned across the gap during tuning.

The light was then focused in the second crystal, which had antireflection coatings but was otherwise identical to the first.  The upconverted flux passed through a BG39 green filter, was focused into a multimode fiber, and was sent to a SPCM-AQR-13-FC single-photon counting module.  Using prism insertion to adjust dispersion, the upconversion rate for $\tau \! = \! 0$ was maximized  (hereafter referred to as `optimum dispersion').  This maximum rate was found to occur for $ 1.5^\circ \,  \mathrm{C}$ less than the axial degenerate-phasematching temperatures of the crystals ($ \approx \! 50^\circ \, \mathrm{C}$), which slightly opens the phase-matched emission and acceptance cones.  In our results, data are the count rate obtained over $6 \, \mathrm{s}$ and, at the close of each data set, the background rate (detector dark count plus stray light, $\approx \! 195  \, \mathrm{s}^{-1}$) was measured over $30 \, \mathrm{s}$ and subtracted.

%2
\begin{figure}
\includegraphics[width=0.48 \textwidth]{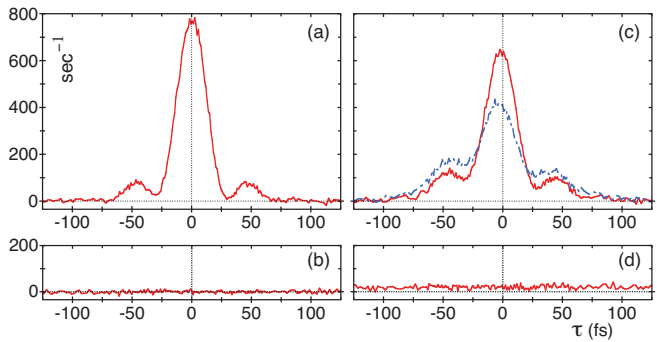}
\caption{Experimental results. (a) Upconversion rate $R(\tau)$ versus signal/idler delay $\tau$ for optimum dispersion.  (b) $R(\tau)$ with upconversion crystal $20^\circ \, \mathrm{C}$ above phasematching temperature.  Other results are for additional signal/idler paths in $6 \, \mathrm{mm}$ of fused silica (c, solid curve), $12 \, \mathrm{mm}$ of fused silica (c, dot-dashed curve), and $37 \, \mathrm{mm}$ of SF10 glass (d), showing increasing broadening of $R(\tau)$.  Data in (c)-(d) have been scaled to correct for Fresnel window losses.}
\end{figure}

In Fig.~2(a) we show results for the upconversion rate $R(\tau)$ with optimum dispersion, where a rate of nearly $ 800  \, \mathrm{s}^{-1}$ appears at $\tau \! = \! 0 \, $.  This peak decays with width (full width at half maximum throughout) $27.9 \, \mathrm{fs}$, and minima and secondary maxima appear near, respectively, $\tau \! = \! \pm \, 30 \, \mathrm{fs}$ and $\pm \, 44 \, \mathrm{fs}$.  For $\left| \tau \right| \! > \! 70 \, \mathrm{fs}$, $R(\tau)$ decays to a noise floor of essentially zero mean.   While statistical error bars are too narrow to be shown clearly, they are consistent with the residual fluctuations seen (error bars of $\pm \, 13  \, \mathrm{s}^{-1}$ near $\tau \! = \! 0 $, and $\pm \, 6  \, \mathrm{s}^{-1}$ in the noise floor, for example).  This low noise floor indicates that random upconversion (i.e., that between unpaired photons) is not significant under the experimental conditions.  As a check on the validity of these data, Fig.~2(b) shows data taken under identical conditions, but with phasematching suppressed in the upconversion crystal by raising its temperature by $20^\circ \, \mathrm{C}$.  Here $R(\tau)$ has fallen to a noise level of essentially zero mean, further supporting that the photons of Fig.~2(a) arose from the desired, phase-matched process.

The consequences of the temporal stretching of the photon wavepackets are shown in Figs.~2(c)-(d).  Returning to the optimum case of Fig.~2(a), windows of fused silica were then inserted just before the fourth prism of Fig.~1, spanning the signal and idler paths, to introduce an identical group delay dispersion  $\phi^{\prime \prime}$ in each mode (taken as the second derivative of accumulated window phase, calculated at degenerate frequency).  In Fig.~2(c) ($6 \, \mathrm{mm}$  thickness of fused silica, $\phi^{\prime \prime}  \! = \! 99 \, \mathrm{fs}^2$), the peak of $R(\tau)$ has fallen to $\approx 640  \, \mathrm{s}^{-1}$ and has widened slightly to $29.4 \, \mathrm{fs}$.  Using a similar window of twice this thickness ($\phi^{\prime \prime}  \! = \! 198 \, \mathrm{fs}^2$), also in Fig.~2(c), the upconversion peak continues to fall ($\approx \! 420  \, \mathrm{s}^{-1}$) and widen ($36.0 \, \mathrm{fs}$ width).  Finally, we have the extreme case of Fig.~2(d), where a window of $37 \, \mathrm{mm}$ of SF10 glass ($\phi^{\prime \prime}  \! = \! 3790 \, \mathrm{fs}^2$) was employed.  A low, featureless result is obtained at a level of $20  \pm 6 \, \mathrm{s}^{-1}$, indicating that $R(\tau)$ has been so broadened that pairwise upconversion is scarcely observable. 

%3
\begin{figure}
\includegraphics[width=0.48 \textwidth]{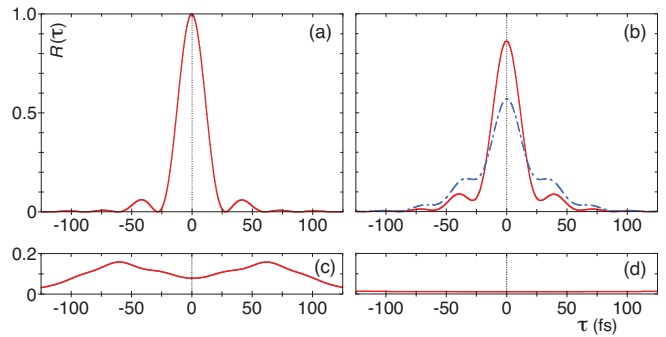}
\caption{Theoretical results. (a) Upconversion rate $R(\tau)$ (arbitrary units) versus signal/idler delay $\tau$ for optimum dispersion.  Also shown is $R(\tau)$ for additional common signal/idler dispersion due to $6 \, \mathrm{mm}$ of fused silica (b, solid curve, $\phi^{\prime \prime}  \! = \! 99 \, \mathrm{fs}^2$), $12 \, \mathrm{mm}$ of fused silica (b, dot-dashed curve, $\phi^{\prime \prime}  \! = \! 198 \, \mathrm{fs}^2$), $5 \, \mathrm{mm}$ of SF10 glass (c, $\phi^{\prime \prime}  \! = \! 513 \, \mathrm{fs}^2$), and $37 \, \mathrm{mm}$ of SF10 glass (d, $\phi^{\prime \prime}  \! = \! 3790 \, \mathrm{fs}^2$).
\label{Fig:Theo}}
\end{figure}

We now compare with theoretical results. Using the experimental parameters, $R(\tau)$ was computed from the squared modulus of the term in the square brackets of Eq.~\ref{E:PsiPUC} with $\phi_s(\omega) \! = \! \phi_i(\omega)$, the integral being done numerically, and the optical system dispersion being calculated exactly. As in the experiment, prism insertion was varied to find optimum dispersion (i.e., maximum $R(0)$); this occurred for a residual system (from first- to second-crystal centers) group delay dispersion of $28 \, \mathrm{fs}^2$, which compensates for quartic dispersion.   In Fig.~3(a), this optimal $R(\tau)$ presents a main peak of width $25.0 \, \mathrm{fs}$, with secondary maxima at $\tau \! = \! \pm \, 42 \, \mathrm{fs}$.  Fig.~3(b) shows $R(\tau)$ for the additional signal/idler dispersion of 6 and $12 \, \mathrm{mm}$ of fused silica, which lowers the peaks and broadens the widths to, respectively,  $26.0 \, \mathrm{fs}$ and $30.9 \, \mathrm{fs}$.  This trend continues with increasing dispersion in Figs.~3(c) and 3(d) to the extent that peaks no longer appear at $\tau \! = \! 0 $.  The comparison with our experimental results is quite good, as may be seen by comparing Figs.~3(a), 3(b), and 3(d) with plots at similar spatial positions in Fig.~2.  The most significant difference is that the calculated $R(\tau)$ is slightly narrower; much of this difference arises because the theoretical single-photon SPDC bandwidth ($130 \, \mathrm{nm}$) is broader than that experimentally measured ($115 \, \mathrm{nm}$), which implies shorter wavepackets in the calculations.

Experimental results that parallel ours, but that have significant differences, have been reported elsewhere\cite{Peer05}.  There, a V-shaped spectral phase mask centered at the degenerate frequency was applied to both photons, producing a relative time shift between the lower- and higher-frequency pair members.  It is readily shown that this amounts to replacing the factor $\exp(-i \, \omega_s \tau)$ in Eq.~\ref{E:PsiPUC} with $\exp(-i  \left|  \omega_s - \omega_p/2 \right|  \tau )$ which, for  a given $S(\omega_s)$, produces a \textit{different} upconversion trace $R(\tau)$.  For example, with a Gaussian  $S(\omega_s)$ our Eq.~\ref{E:PsiPUC} predicts a Gaussian $R(\tau)$, while the approach of Ref.~\onlinecite{Peer05} predicts a curve 1.7 times wider, with tails similar to a Lorentzian.  Moreover, it is readily shown that the approach of Ref.~\onlinecite{Peer05} predicts skewness in curves analogous to Fig.~3(b), since dispersion produces time shifts between the lower- and higher-frequency pair members (although we do not reproduce these calculations here).  In contrast, our experiment instead relies on \textit{spatial} separation to define the signal and idler modes, and the spectral content of each mode is left intact.  Thus, while our method produces a pure time shift between the two modes, the approach of Ref. ~\onlinecite{Peer05} implies a temporal re-shaping of the single photons.

It is also notable that upconversion of single photons with a classical laser beam has been demonstrated\cite{LaserUp}, and an extension of this approach has been reported in which each member of a given signal/idler pair upconverts with the same ultrafast laser pulse\cite{Kuzucu08}.  What is obtained, after averaging, is the two-dimensional histogram of signal and idler arrival times with respect to the laser pulses, with resolution determined by the pulse width ($150 \, \mathrm{fs}$)\cite{Kuzucu08}.  In comparison, our method is intrinsically only one-dimensional in time, although it has at least 2 orders of magnitude more resolution, depending simply on mirror positioning accuracy.  Finally, it is significant that our approach is highly sensitive to group delay dispersion of the photon pairs while, in the similar case of a monochromatic SPDC pump, the Hong-Ou-Mandel interferometer is not\cite{Stein92}.  In fact, in the phase term of Eq.~(\ref{E:Somega}), the phase factor $\exp{(i \overline{\Delta k}_z L/2)} \exp{\{i[\phi_s(\omega_s)+\phi_i({\overline{\omega}_i})]\}} $ is that associated with the dispersed two-photon state \textit{itself}, and thus our approach provides a fairly direct, high-sensitivity means of probing the consequences of its dispersion.  Due to slow detector response, observing group delay dispersion effects through direct detection of signal and idler arrival times has required  $\approx \! 10^5$ times more dispersion than we easily see, for example, in the cases of Figs.~2(c) and 3(b)\cite{Valencia02}.  Based on the sensitivity of our calculations, we estimate that $\approx \! 10 \, \mathrm{fs}^2$ of group delay dispersion could be observed with our method (in either the signal or idler mode, or distributed in both, as is clear from the phase terms of Eq.~(\ref{E:Somega})).  However, an experimental demonstration of this sensitivity is beyond the scope of our work.

In conclusion, we have reported results in which entangled photon pairs are upconverted with a controlled time delay introduced between the signal and idler photons.  We have thus demonstrated an ultrafast, nonclassical correlation between the signal and idler waves, even for our case of a continuous-wave, single-mode SPDC pump.  Our theoretical analysis has developed an expression for the upconverted amplitude that is strongly dependent on the dispersion of the signal and idler modes.  In experiments, the signal photons have been spatially separated from the idlers, delayed with the fs-level resolution inherent in a piezoelectric stage, and then recombined with the idlers for upconversion.   Close agreement has been demonstrated between the observed and predicted upconversion rates as a function of signal/idler time delay. We believe that this technique is a valuable tool for the characterization of entangled photon pairs at the fs level. 

\begin{acknowledgments}
This research was supported in part by CONACYT (Mexico) under Grant 49570-F.
\end{acknowledgments}

\end{document}